\documentclass[prl,amsmath,amsfont,amssymb,twocolumn,showpacs]{revtex4-1}
\usepackage{graphicx}
\usepackage{color}
\usepackage[english]{babel}
\usepackage{amsmath}
\usepackage{bm}
\usepackage{graphicx}
\usepackage{setspace}
\usepackage{pgfplots}
\usepackage{comment}
\usepackage{subfigure}
\begin{document}
\def\beq{\begin{equation}}
\def\eeq{\end{equation}}
\def\bea{\begin{align}}
\def\eea{\end{align}}
\def\ba{{\bf a}}
\def\brho{{\boldsymbol \rho}}
\def\bk{{\bf k}}
\def\bp{{\bf p}}
\def\bq{{\bf q}}
\def\br{{\bf r}}
\def\bv{{\bf v}}
\def\bx{{\bf x}}
\def\bz{{\bf z}}
\def\bG{{\bf G}}
\def\bC{{\bf C}}
\def\bF{{\bf F}}
\def\bJ{{\bf J}}
\def\bP{{\bf P}}
\def\la{\langle}
\def\ra{\rangle}
\def\calH{\mathcal{H}}
\def\calU{\mathcal{U}}
\def\calD{\mathcal{D}}
\def\calC{\mathcal{C}}
\def\calK{\mathcal{K}}
\def\calP{\mathcal{P}}
\def\calV{\mathcal{V}}
\def\calE{\mathcal{E}}
\def\p{\hat {\psi}} 
\def\pd{\hat {\psi}^{\dag}}
\def\grad{\mbox{\boldmath $\nabla$}}
\def\Tr{{\rm Tr}}
\def\e{\epsilon}
\def\ve{\varepsilon}
\def\pa{\partial}
\def\nn{\nonumber}
\def\t{\tau}
\def\kbar{\bar {k}}
\def\rbar{\bar {r}}
\def\nbar{\bar {n}}

\title{Coexistence of  density wave and superfluid order in a 
dipolar Fermi gas}
\author{Zhigang Wu, Jens K. Block and Georg M. Bruun}
\affiliation{Department of Physics and Astronomy, Aarhus University, DK-8000 Aarhus C, Denmark}
\date{\today}
\begin{abstract}
We analyse the coexistence of superfluid and density wave (stripe) order in a  quasi-two-dimensional gas of dipolar fermions aligned by an external field. 
Remarkably, the anisotropic nature of the dipolar interaction allows for such a coexistence in a large region of the zero temperature phase diagram. 
In this region, the repulsive part of the interaction drives the stripe formation and the attractive part induces the pairing, resulting in a 
supersolid with $p$-wave Cooper pairs aligned along the stripes. From a momentum space perspective, the stability of the supersolid phase is due to the fact that the stripe order renders the Fermi surface only partially gapped, leaving gapless regions that are most important for $p$-wave pairing.  
We finally demonstrate how this supersolid phase can be detected in time-of-flight experiments.  
\end{abstract}
\maketitle
Since the prediction of superfluidity in solid helium several decades ago~\cite{Andreev, Leggett, Chester,Boninsegni}, the intriguing 
possibility of coexisting diagonal (density) and off-diagonal (superfluid) 
order forming a supersolid has been subject to intense investigations. However, the supersolid phase has not been observed unequivocally 
as interpretations of state-of-the-art experiments in helium are still debated~\cite{Kim,Balibar}. 
The recent experiments on cold dipolar gases~\cite{Ospelkaus,Gunton,Heo,Wu,Lev,Schulze,Tung,Repp}, may 
finally allow an observation of this conceptually important phase. Supersolidity has been predicted to exist for dipolar bosons in an optical lattice~\cite{Capogrosso,Pollet,Cinti}, dipolar bosons with three-body forces~\cite{Lu}, and spinor Bose condensates with spin-orbit coupling~\cite{Li1,Li2}. For Fermions however, relevant studies are fewer and limited to the case of an optical lattice~\cite{Gadsbolle,Zeng}. In this paper, we  expand  the scope of study concerning  supersolidity of dipolar Fermi gases and show that a supersolid phase is in fact the ground state in a large region of the phase diagram of a two-dimensional (2D) Fermi gas of dipoles aligned by an external field. 

\begin{center}
\begin{figure}[ht]
 \def\svgwidth{0.68\linewidth}
\begingroup%
  \makeatletter%
  \providecommand\color[2][]{%
    \errmessage{(Inkscape) Color is used for the text in Inkscape, but the package 'color.sty' is not loaded}%
    \renewcommand\color[2][]{}%
  }%
  \providecommand\transparent[1]{%
    \errmessage{(Inkscape) Transparency is used (non-zero) for the text in Inkscape, but the package 'transparent.sty' is not loaded}%
    \renewcommand\transparent[1]{}%
  }%
  \providecommand\rotatebox[2]{#2}%
  \ifx\svgwidth\undefined%
    \setlength{\unitlength}{865.2bp}%
    \ifx\svgscale\undefined%
      \relax%
    \else%
      \setlength{\unitlength}{\unitlength * \real{\svgscale}}%
    \fi%
  \else%
    \setlength{\unitlength}{\svgwidth}%
  \fi%
  \global\let\svgwidth\undefined%
  \global\let\svgscale\undefined%
  \makeatother%
  \begin{picture}(1,0.74898867)%
    \put(0,0){\includegraphics[width=\unitlength]{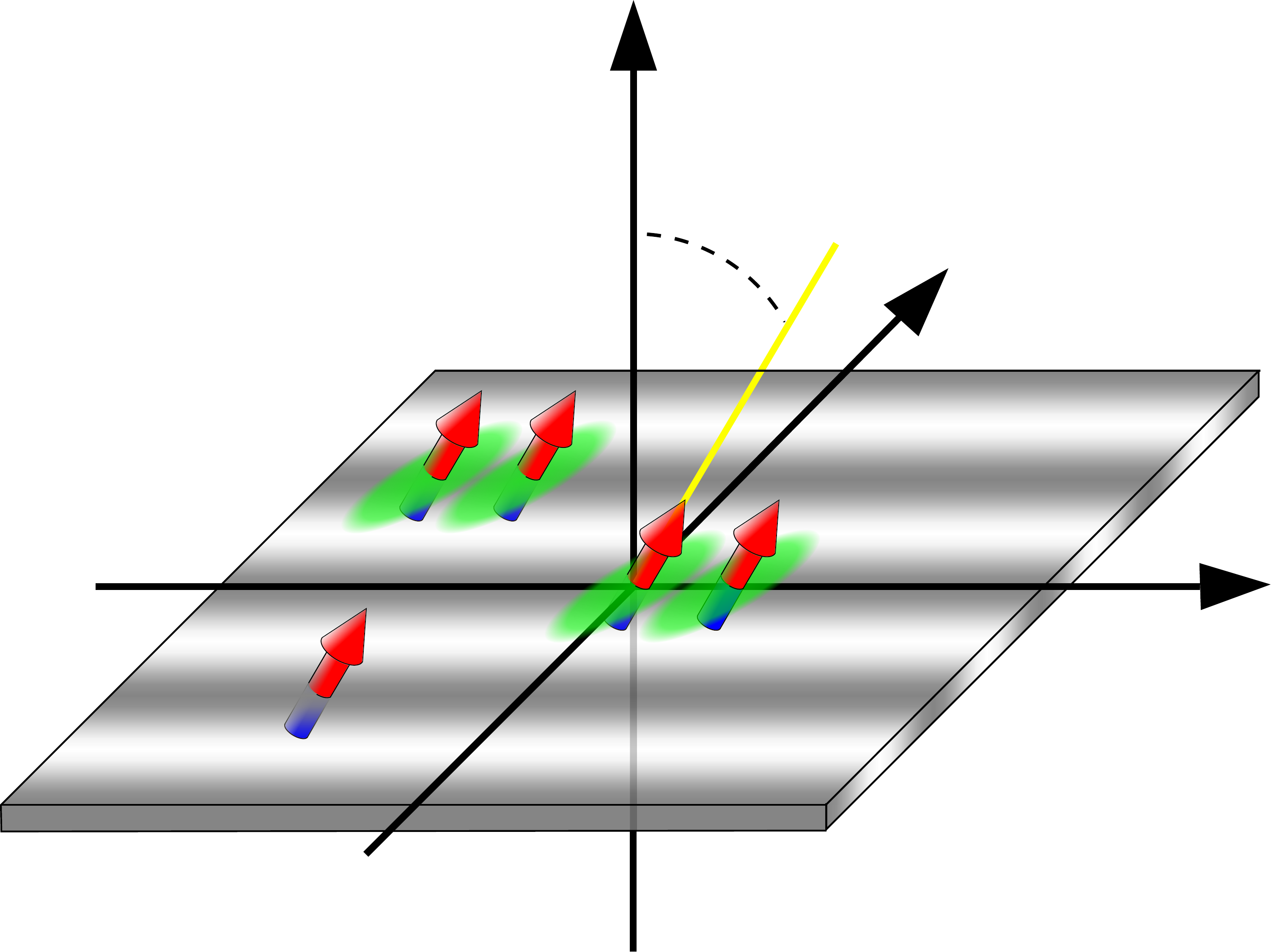}}%
    \put(0.64942817,0.58655036){\color[rgb]{0,0,0}\makebox(0,0)[lb]{\smash{$\mathbf{E}$}}}%
    \put(0.87847367,0.22792689){\color[rgb]{0,0,0}\makebox(0,0)[lb]{\smash{x}}}%
    \put(0.75051751,0.48038088){\color[rgb]{0,0,0}\makebox(0,0)[lb]{\smash{y}}}%
    \put(0.4360548,0.64637805){\color[rgb]{0,0,0}\makebox(0,0)[lb]{\smash{z}}}%
    \put(0.5518844,0.56452813){\color[rgb]{0,0,0}\makebox(0,0)[lb]{\smash{$\Theta$}}}%
  \end{picture}%
\endgroup%
\caption{Illustration of a 2D Fermi gas with  dipoles aligned by an external field ${\bf E}$ in the supersolid phase. Stripes with high density  are indicated with a 
dark color, and the $p$-wave nature of the pair wave function is indicated by green regions.  }
\label{SetupFig}
\end{figure}
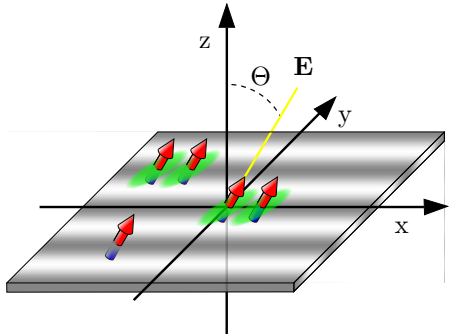
\end{center}

We consider spinless Fermions with a dipole moment $\bf d$ at zero temperature, confined in the $xy$ plane by a  harmonic trapping potential $V_{\rm tr}(\br) = m\omega_z^2z^2/2$ along the $z$-direction. We take  $\omega_z \gg \e_F^{0}$ ($\hbar = 1$), where $\e_F^{0}={k^{0}_F}^2/2m$ is the Fermi energy of a 2D non-interacting gas with areal density $n_0$ and $k^0_F = \sqrt{4\pi n_0}$.  In this limit, the Fermions are ``frozen" in the harmonic oscillator ground state in the $z$ direction and the system is  effectively 2D. The dipole moments are  aligned by an external field ${\bf E}$ into a direction which is perpendicular to the $y$-axis and forms an angle $\Theta$ with respect to the $z$-axis as illustrated in Fig.\ \ref{SetupFig}. 
The interaction between the dipoles is $V_{\rm d}(\br)  = D^2(1-3 \cos^2\theta_{rd})/r^3$ where $D^2 = d^2/4\pi\ve_0$ for electric dipoles, and $\theta_{rd}$ is the angle between the relative displacement vector of the two dipoles $\br=(\brho,z)$ and the dipole moment $\bf d$. For $\omega_z \gg \e_F^{0}$,
the Fourier transform of the effective 2D interaction is given by 
(up to an irrelevant constant term)~\cite{Fischer}
\begin{align}
V(\bq) &= -2\pi D^2F(q)\xi(\Theta,\varphi).
\label{V2Df1}
\end{align}
Here $F(q) = q\exp[(qw)^2/2]{\rm erfc}(qw/\sqrt{2})$ and $\xi(\Theta,\varphi)=\cos^2\Theta-\sin^2\Theta\cos^2\varphi$, where $w=\sqrt{1/m\omega_z}$ is the trapping length in $z$-direction and $\varphi$ is the polar angle of $\bq$. We note that $F(q)$ saturates in the limit of large $q$. This of course is not physical since any true molecular potential has a strong repulsive core which effectively provides a high momentum cut-off for the potential. Such a cut-off can also be introduced by considering the two-body  scattering problem as we will discuss below. With a cut-off in mind, we can use the fact that the 2D limit $\omega_z \gg \e_F^{0}$ is equivalent to $k^0_{F}w\ll 1$, and make the approximation $F(q)\simeq q + {\mathcal O}(qw)$ in (\ref{V2Df1}). 

The strength of this  interaction is measured by the ratio of the typical interaction and kinetic energy $g = 4mD^2k^0_{F}/3\pi$. In addition to $g$, the system is characterised by the dipole tilting angle $\Theta$, which controls the degree of anisotropy of the interaction in the $xy$-plane. For weak to moderate interaction strengths and small tilting angles, the system is well described by Landau Fermi liquid theory~\cite{Baranov2,Lu}. For $\Theta>0$, 
  the repulsion between two dipoles in the $xy$ plane is strongest when their relative displacement vector is along the $y$ direction and weakest along the $x$ direction. The anisotropy is predicted to give rise to stripe formation for interaction strengths beyond a critical value $g_c(\Theta)$~\cite{Block1,Baranov2,Babadi,Parish,Block2}. In this phase the dipoles form stripes parallel to the $x$-axis to minimise the repulsion, corresponding to a density modulation with wave vector $\bq_c=q_c \hat {\bf y}$ as illustrated in Fig.~\ref{SetupFig}. As $\Theta$ increases the dipolar interaction eventually becomes partially attractive. A  Fermi liquid to $p$-wave superfluid phase transition is predicted
   to occur for tilting angles greater than a critical angle $\Theta_s\simeq \arcsin(2/3)\simeq0.23\pi$, due to the attractive part of the dipolar interaction~\cite{Bruun}. 

\begin{center}
\begin{figure}[ht]
 \includegraphics[width=0.68\linewidth]{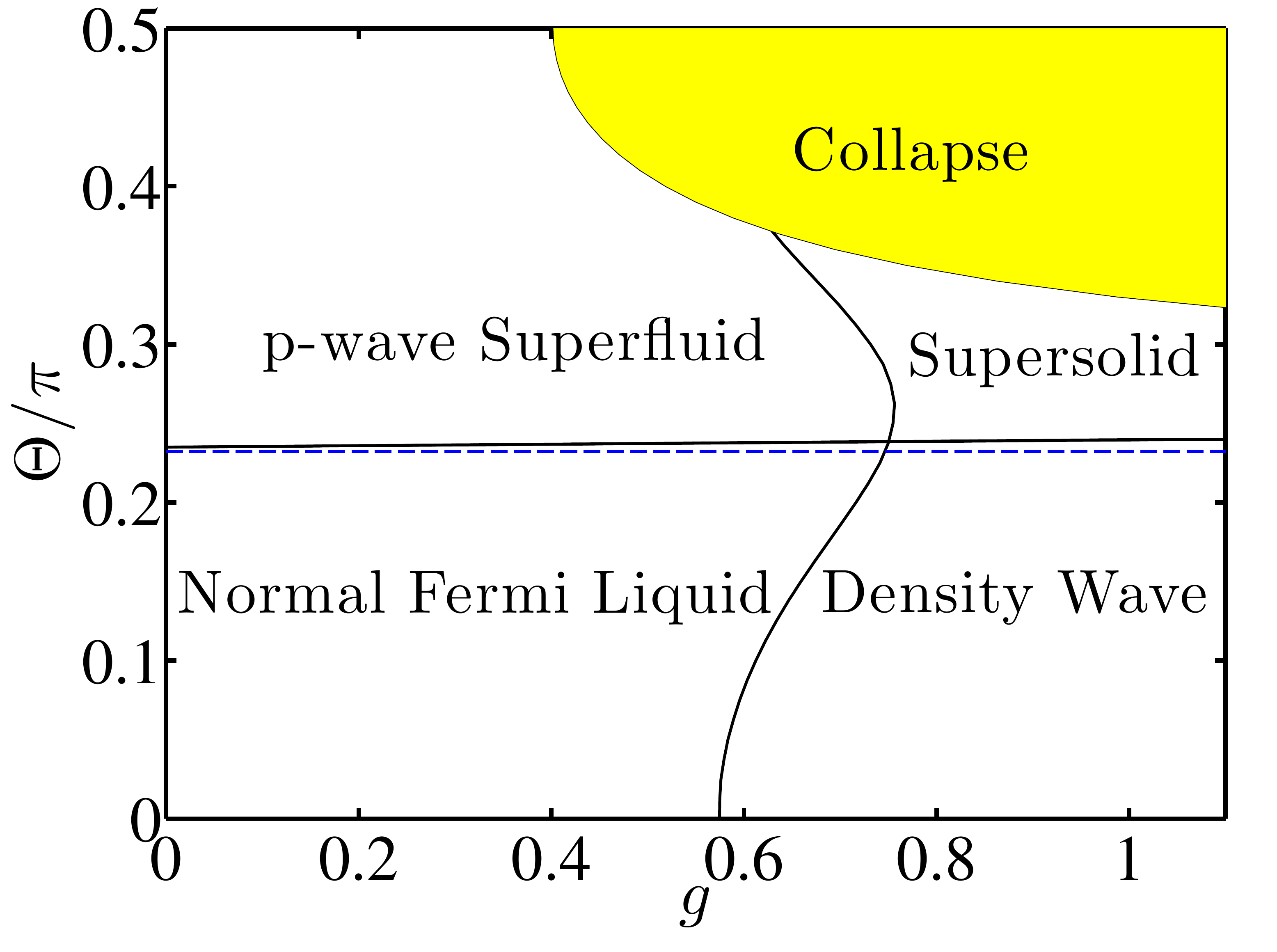}
\caption{Mean-field phase diagram of 2D dipolar Fermi gas. The dashed line is $\Theta = \arcsin(2/3)$ and the solid line just above it is obtained from a more accurate calculation.}
\label{PhaseFig}
\end{figure}
\end{center}

The zero temperature mean-field phase diagram shown in Fig.\ \ref{PhaseFig} summarises the discussion given above. In this phase diagram the critical coupling strength $g_c(\Theta)$ for stripe formation is obtained from a Hartree-Fock (HF) calculation~\cite{Block1,Block2} and the normal Fermi liquid to superfluid transition critical angle is determined by  BCS theory (see below). Although each of these three phases have been studied extensively, 
an interesting question remains unaddressed: what  is the nature of the ground state in the region of the phase diagram where the superfluid and stripe phases overlap? In this paper, we provide an answer to this question. We demonstrate that the system in the density wave phase eventually becomes unstable towards pairing as the tilting angle increases and the interaction becomes more attractive. Importantly, the resulting superfluid order does not exclude the stripe order, thus making the system a supersolid as understood in the sense mentioned above.

\begin{center}
\begin{figure}[ht]
\subfigure{\includegraphics[width=0.49\linewidth]{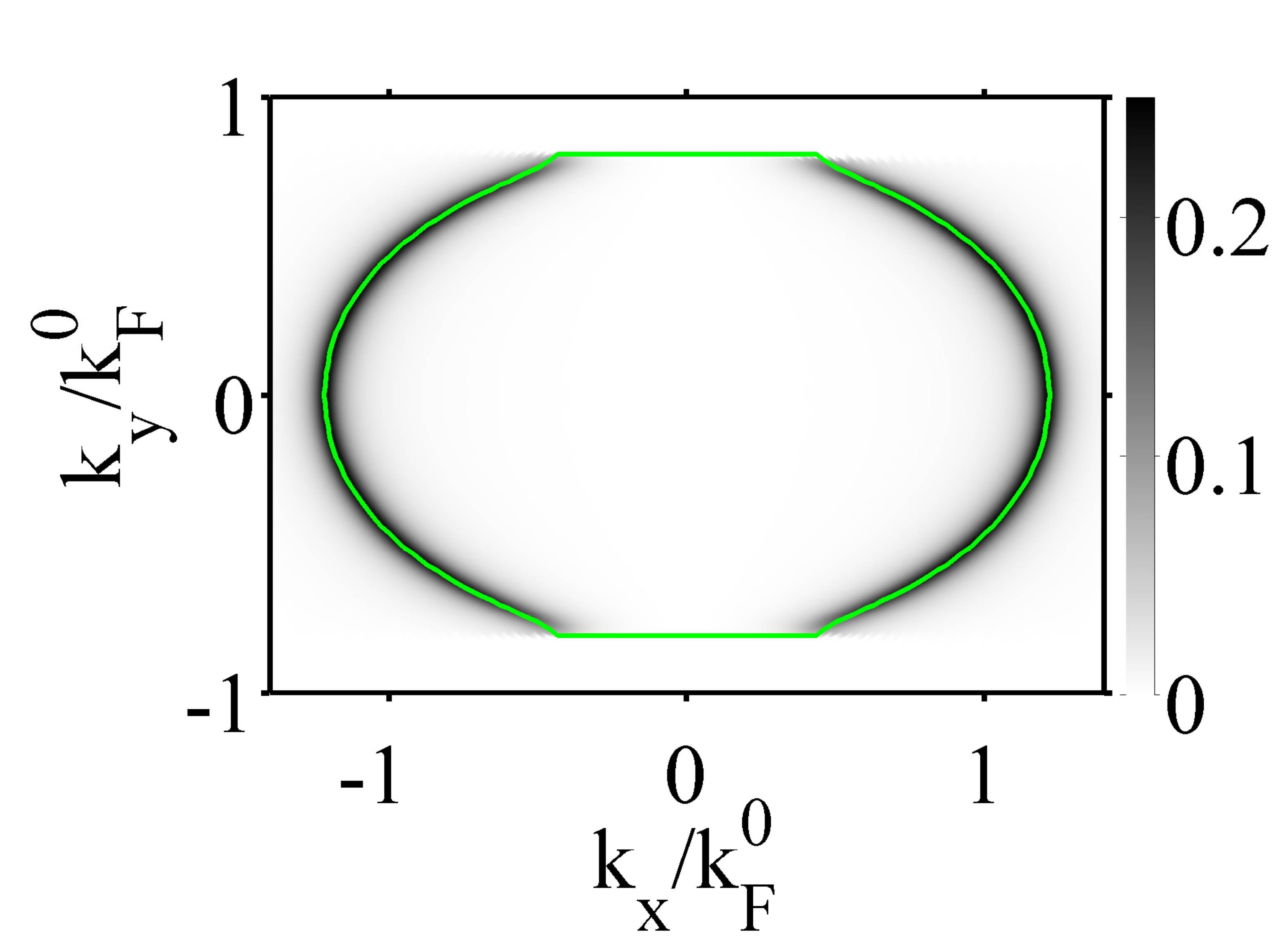}}
\subfigure{\includegraphics[width=0.49\linewidth]{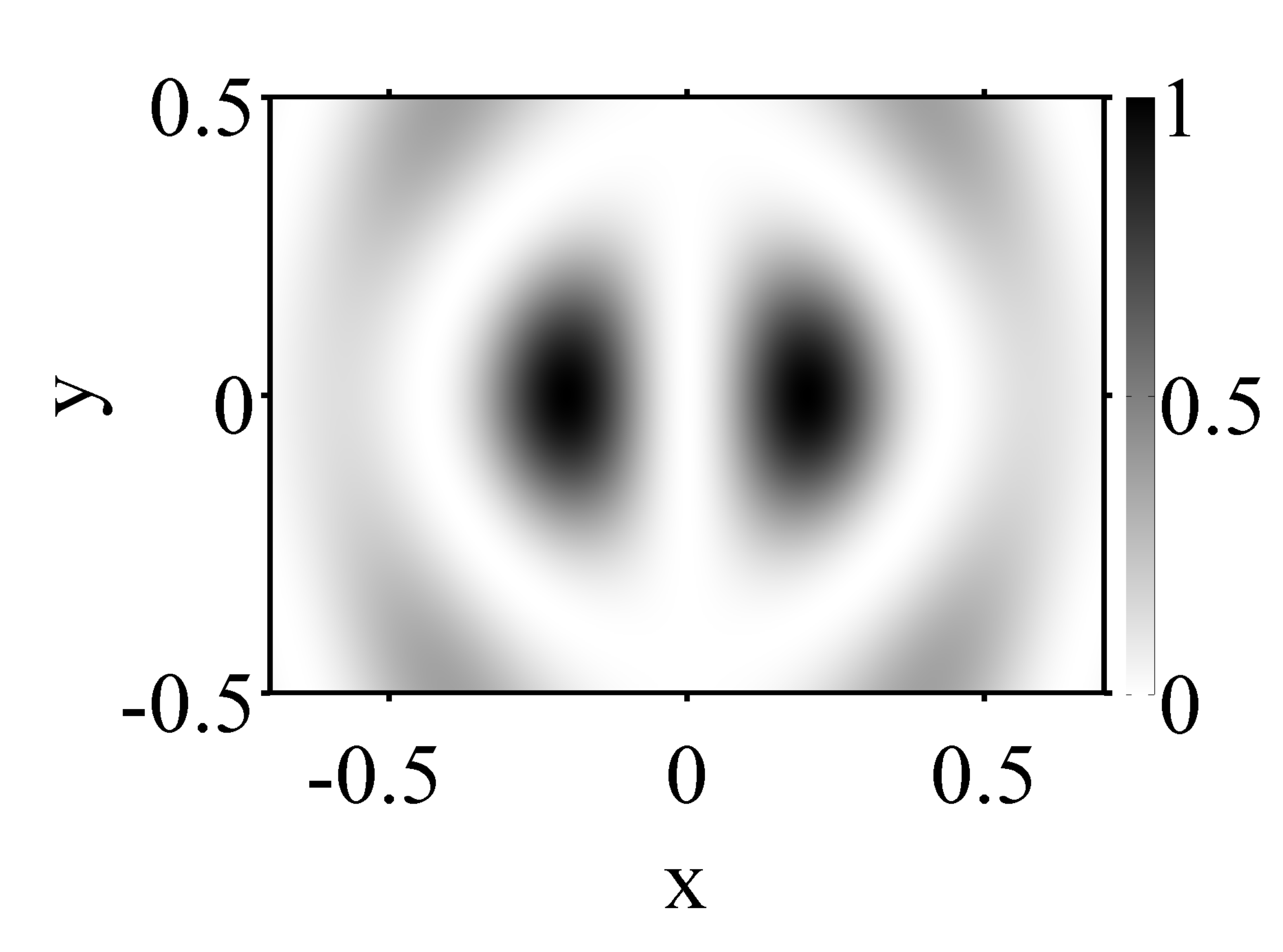}}
\caption{Left: The pair correlation function for $g=1.0$ and $\Theta = 0.28\pi$. The Fermi surface $k_F(\phi)$ in the stripe phase is shown by a green line. 
Right: $|\psi_{pair}(\brho,-\brho)|^2$ (normalised to the maximum value) as a function of $\brho$ for the same parameters. Here the lengths are in units of $2\pi/q_c$.}
\label{pair-corr}
\end{figure}
\end{center}

We use self-consistent HF theory to analyse the stripe phase, as it is the only theory so far that allows us to determine quantitatively properties of this phase. 
In the stripe phase, the translational symmetry is broken in the $y$ direction, and $\la\hat c_\bk\hat c^\dag_{\bk\pm\bq_c} \ra\neq 0$, where $\hat c_{\bk}$ annihilates a dipole with momentum $\bk$. This yields a modulated density $n(\brho)=n_0+n_1\cos(\bq_c\cdot\brho)$ where $n_1$ is the stripe order parameter. As described in detail in Ref.\ \cite{Block2}, the resulting mean-field Hamiltonian can be diagonalised by an unitary transformation $\hat \gamma_{j\bar\bk} = \sum_\bG U_{j,\bar\bk+\bG}\hat c_{\bar\bk + \bG}$, where $\bar\bk$ is restricted to the first Brillouin zone (BZ) $-q_c/2 <\bar\bk\cdot\hat\bq_c \leq q_c/2$, $\bG = l \bq_c, l =0,\pm 1,\cdots$ are the reciprocal lattice vectors, and $j$ is a band index. The diagonalised  Hamiltonian takes the form 
$
\hat \calH_{\rm MF}=\sum_{j\bar\bk} \varepsilon_{j\bar\bk}\hat \gamma_{j\bar\bk}^\dag\hat \gamma_{j\bar\bk}
$,
where $\hat \gamma_{j\bar\bk}$ annihilates a quasiparticle $\psi_{j\bar\bk}(\brho)$ with energy $\varepsilon_{j\bar\bk}$. The most salient outcome of the HF analysis
is that the Fermi surface $k_F(\phi)$ contains gapped regions around $\phi=\pm\pi/2$, as well as gapless regions around $\phi=0$ and $\phi=\pi$, see Fig.~\ref{pair-corr}. 
The gapped regions are  a manifestation of the stripe order, and the gap magnitude is roughly proportional to $\e_F^0 n_1/n_0$. 

The key fact for the present purpose is that the stripe order still leaves gapless regions on the Fermi surface, which opens up the intriguing possibility of superfluid pairing. To explore this, we use BCS theory with the Hamiltonian $\hat\calH_{\rm BCS}=\hat \calH_{\rm MF}+\hat \calH_{\rm P}$. Here,   
 \begin{align}
\hat \calH_{\rm P} = \sum_{{jj' 
 \bar\bk\bar\bk'}}\frac{\calV_{j'j}(\bar\bk',-\bar\bk)}2\la \hat \gamma^\dag_{j'\bar\bk'}\hat \gamma_{j',-\bar\bk'}^\dag\ra\hat \gamma_{j\bar\bk}\hat \gamma_{j,-\bar\bk} + \rm{h.c.} \nn
\end{align}
describes pairing between the time-reversed states, where $\calV_{j'j}(\bar\bk',-\bar\bk) $
is the interaction between the quasiparticles~\cite{SM}. 
To derive a gap equation that is amenable to a partial wave expansion, we switch to the ``extended zone scheme", whereby a single particle state $\psi_{j\bar\bk}(\brho)$ in the $j$'th band in the first BZ is mapped onto a state $\psi_{\bk}(\brho)$ in the $j$'th BZ in the standard way~\cite{SM}, where the vector $\bk$ is now unrestricted.
The effective pairing interaction $\calV_{j'j}(\bar\bk',-\bar\bk) $ shall be denoted by  $\calV(\bk,-\bk')$.
Pairing between time-reversed quasiparticles gives rise to the gap parameter 
$
\Delta_{\bk}  \equiv \sum_{\bk'} \calV(\bk,-\bk')\la\hat \gamma_{-\bk'}\hat \gamma_{\bk'} \ra
$, which satisfies the gap equation   
\beq
\Delta_\bk = -\int\frac{d\bk'}{(2\pi)^2}\calV(\bk,-\bk')\Delta_{\bk'}\left (\frac{1}{2E_{\bk'}} -\frac{\calP}{2\xi_{\bk'}}\right ).
\label{gapeq2}
\eeq
Here $\xi_\bk = \varepsilon_\bk-\mu$ and $E_\bk=\sqrt{\xi_{\bk}^2+|\Delta_\bk|^2}$, where the chemical potential $\mu$ is approximated by the value in the stripe phase.
The Cauchy principal value term $\calP/2\xi_{\bk'}$ in (\ref{gapeq2}) renders the gap equation well defined with no need for a high momentum cut-off. Such a term can be introduced by renormalizing the gap equation in terms of scattering amplitude of two dipoles in a vacuum~\cite{Levinsen,Cooper,Baranov2}. In the absence of experimental data for dipole-dipole scattering in 2D, one can simply regard it as a specific procedure to provide a cut-off.  
 
To solve the gap equation, we expand the gap parameter as $\Delta_\bk = \sum_{n=1}^\prime \Delta_n(k) \cos n\phi$ where $\sum'$ 
 restricts the summation to odd indices,
since  $\Delta_{-\bk} = -\Delta_\bk$ for spinless Fermions. 
 A more general  expansion contains both $\cos n\phi$ and $\sin n\phi$ terms. However, the $\cos n\phi$ terms are favoured by the attractive part of the potential and the gap parameter given by the previous expression maximises the  pairing~\cite{SM}. Using the expansion in (\ref{gapeq2}) we obtain a system of equations 
 \beq
\Delta_n(k) = \sideset{}{'}\sum_{n'=1}^\infty \int_0^\infty  dk'  \calK_{nn'}(k,k')\Delta_{n'}(k'),
\label{geqdw2}
\eeq
where 
\begin{align}
\calK_{nn'}(k,k') &= -\frac{1}{8\pi^2}\sideset{}{'}\sum_{l=1}^\infty k'\calV^{cc}_{nl}(k,k') \nn \\
\times&\int_0^{2\pi} \,{d\phi'}\cos l\phi'\cos n'\phi'\left (\frac{1}{E_{\bk'}} -\frac{\calP}{\xi_{\bk'}}\right )
\label{calK}
\end{align}
and 
\begin{align}
\calV^{cc}_{nn'}(k,k') =\iint\limits_{0}^{\quad\,\, 2\pi}\,\frac{d\phi}\pi \frac{d\phi'}{\pi} \cos n \phi \cos n' \phi' \calV(\bk,-\bk').
\label{calVcc}
\end{align}
Equations (\ref{geqdw2})-(\ref{calK}) with (\ref{calVcc}) are 
the fundamental equations to be solved numerically, and they form the basis of the results presented in the rest of this paper.

Before we turn to fully numerical solutions of (\ref{geqdw2}), it is important to understand under what conditions the gap equation admits a solution. To do so, we shall first examine the Fourier components $\calV^{cc}_{nn'}(k,k')$.  As shown in the Supplementary Material~\cite{SM}, these Fourier components calculated numerically from (\ref{calVcc}) differ very little from those between the bare particles, which are obtained by replacing $\calV(\bk,-\bk')$ in (\ref{calVcc}) by the bare interaction $V(\bk-\bk')$. The latter components, denoted by $V^{cc}_{nn'}(k,k')$, can be determined analytically and obey the selection rule $V^{cc}_{nn'}(k,k')\neq 0$ only if $n'=n,n\pm 2$. In addition, the lowest component $V^{cc}_{11}(k,k')$ is in general
dominant over the higher components.
 We find that the Fourier components $\calV^{cc}_{nn'}(k,k')$ possess all the above properties to a very good approximation. The agreement between $\calV^{cc}_{nn'}(k,k')$ and $V^{cc}_{nn'}(k,k')$ holds even deep into the stripe phase, which seems initially surprising since a large stripe amplitude gives rise to extended gapped regions around the Fermi surface. The reason is that the quasiparticle interaction $\calV(\bk,-\bk')$ is altered from $V(\bk-\bk')$
  only in the gapped regions centred at $\phi=\pm\pi/2$, 
  which are precisely the regions of integration in (\ref{calVcc}) suppressed by the $\cos n\phi\cos n'\phi'$ factor. In light of earlier work on the superfluid transition~\cite{Bruun}, the fact that the quasiparticle pairing interaction is approximately the same as that between the bare particles strongly suggests that pairing in the stripe phase is possible, provided that the Fermi surface is not fully gapped. 
 
From these results it can be shown that the dominant component of the gap parameter 
 is in fact the first harmonic. Thus the simplest approximation is the momentum-independent ansatz $\Delta_{\mathbf k}\simeq\Delta_1\cos\phi$. Since  
the integrand in (\ref{geqdw2}) is peaked around the partially gapped Fermi surface, a good estimate for when the gap equation admits  a finite 
solution simply follows from the requirement that the effective $p$-wave interaction in the vicinity of the Fermi surface is attractive, i.e.,
\beq
\calV^{cc}_{11}\equiv\calV^{cc}_{11}(k_F(0),k_F(0)) \simeq \frac{4\pi g}{m}\big(1-\frac94\sin^2\Theta  \big ) < 0,
\eeq
where $k_F(0)$ is the Fermi wave vector at $\phi=0$. The critical angle is therefore  $\Theta_s = \arcsin(2/3)$, which is the same as that obtained for a normal Fermi liquid to superfluid transition at the same level of approximation~\cite{Bruun}. 

\begin{center}
\begin{figure}[ht]
\subfigure{\includegraphics[width=0.49\linewidth]{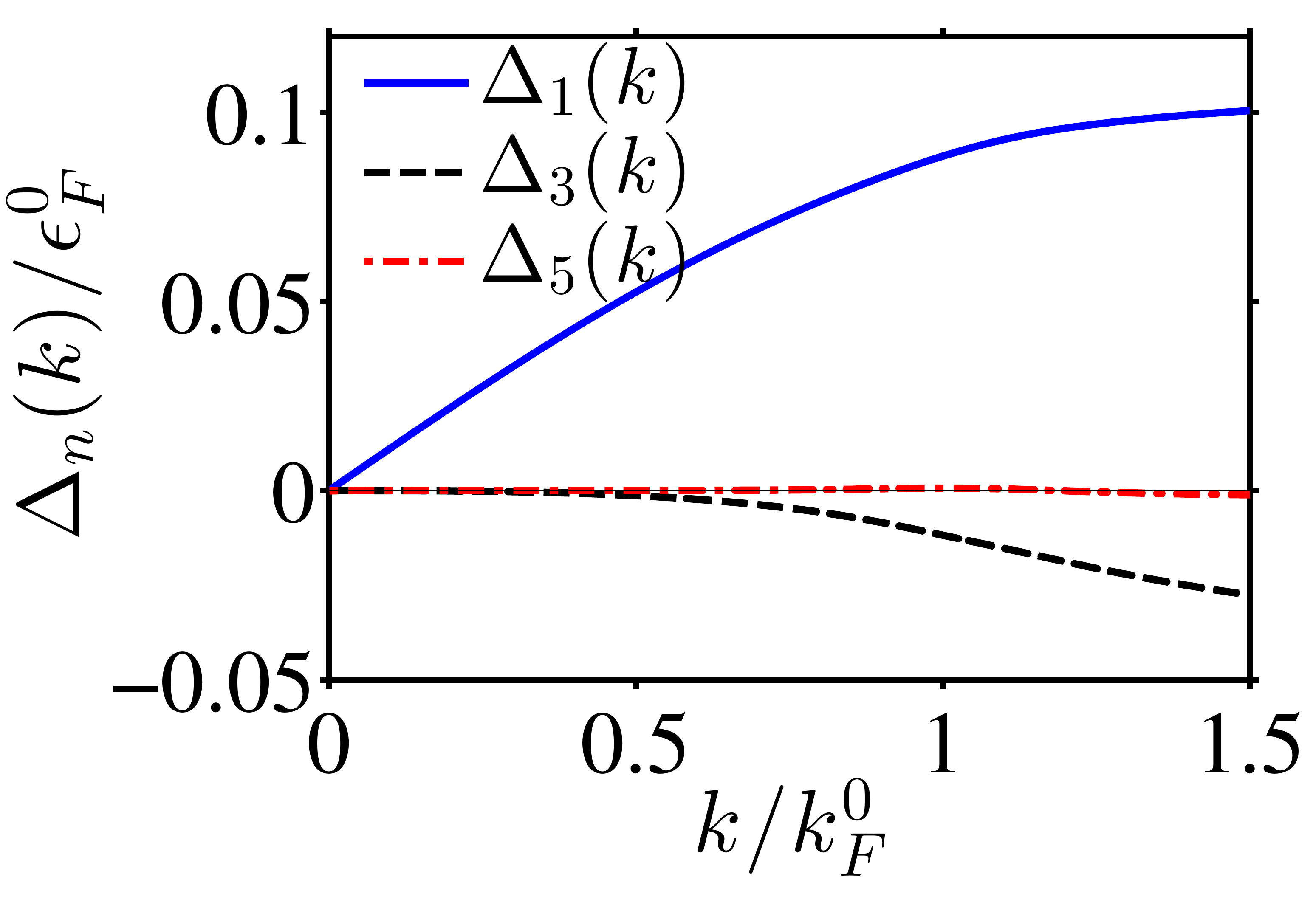}}
\subfigure{\includegraphics[width=0.49\linewidth]{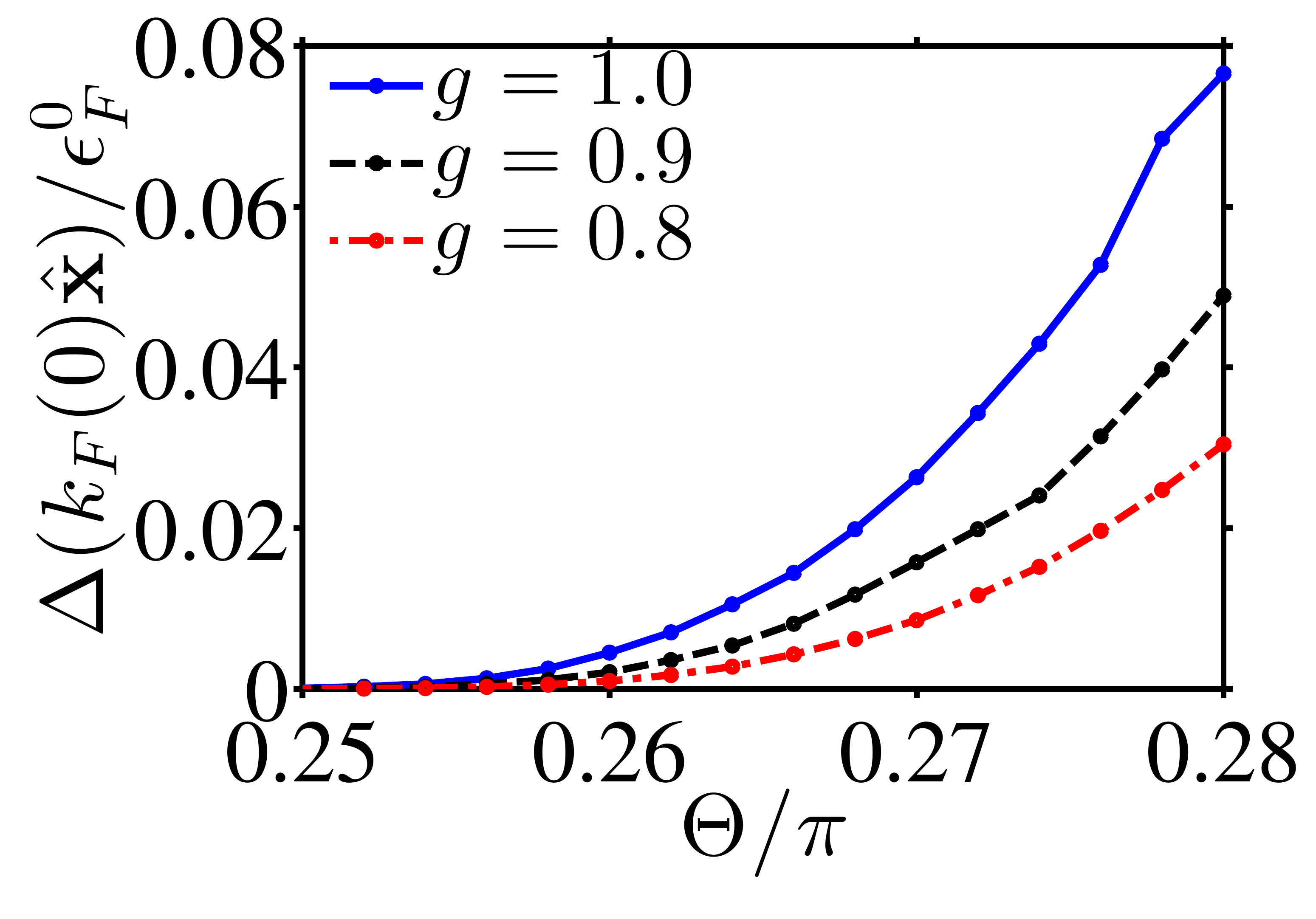}}
\caption{Left: Amplitudes of the gap parameter as a function of $k/k_F^0$ for $g = 1$ and $\Theta = 0.28\pi$. Right: The gap parameter at the tip of Fermi surface as a function of tilting angle $\Theta$.}
\label{gappa}
\end{figure}
\end{center}

With an estimate of the critical angle, we now solve the gap equation self-consistently  including higher harmonics and retaining the full momentum 
dependence of $\Delta_n(k)$. The quasiparticle energies $\xi_\bk$ and 
the effective interactions are calculated from the HF theory for the stripe phase and are then used as input to the gap equation (\ref{geqdw2})-(\ref{calK}). This approach assumes that the pairing has a negligible effect on the stripes, which we will demonstrate is correct. 
As an example of the calculations, we show in Fig.~\ref{gappa} (left) the amplitudes of the first three partial wave components of the gap parameter for $g = 1$ and $\Theta = 0.28\pi$. For these parameters, the system  is deep in the stripe phase with $n_1/n_0 \simeq 0.26$ in the absence of pairing.  
We see that the pairing occurs dominantly in the $p$-wave channel $\cos \phi$, 
but also has a noticeable $f$-wave ($\cos 3\phi$) component; all the higher partial wave components are completely negligible. This feature is in fact typical of solutions to the gap equation. With the solutions of the gap equation we can further determine the redistribution of the quasiparticles and hence, the change of the stripe amplitude as a result of pairing. We find very small 
relative changes in the stripe amplitude in all our calculations. This demonstrates that our approach is consistent and that the stripe and superfluid orders can 
indeed coexist forming a type of supersolid. Figure \ref{gappa} (right) depicts the gap parameter at the tip of the Fermi surface, 
$\Delta(k_F(0)\hat \bx)$, as a function of $\Theta$ for various values of $g$. For negative but small effective $p$-wave interaction, the behaviour of $\Delta(k_F(0)\hat \bx)$ is well 
described by the weak pairing approximation $\Delta(k_F(0)\hat \bx)/\e_F^0 \sim \exp\left (4\pi/m\calV^{cc}_{11}\right )$, which follows from the ansatz mentioned earlier. 

\begin{center}
\begin{figure}[ht]
\includegraphics[width=0.49\linewidth]{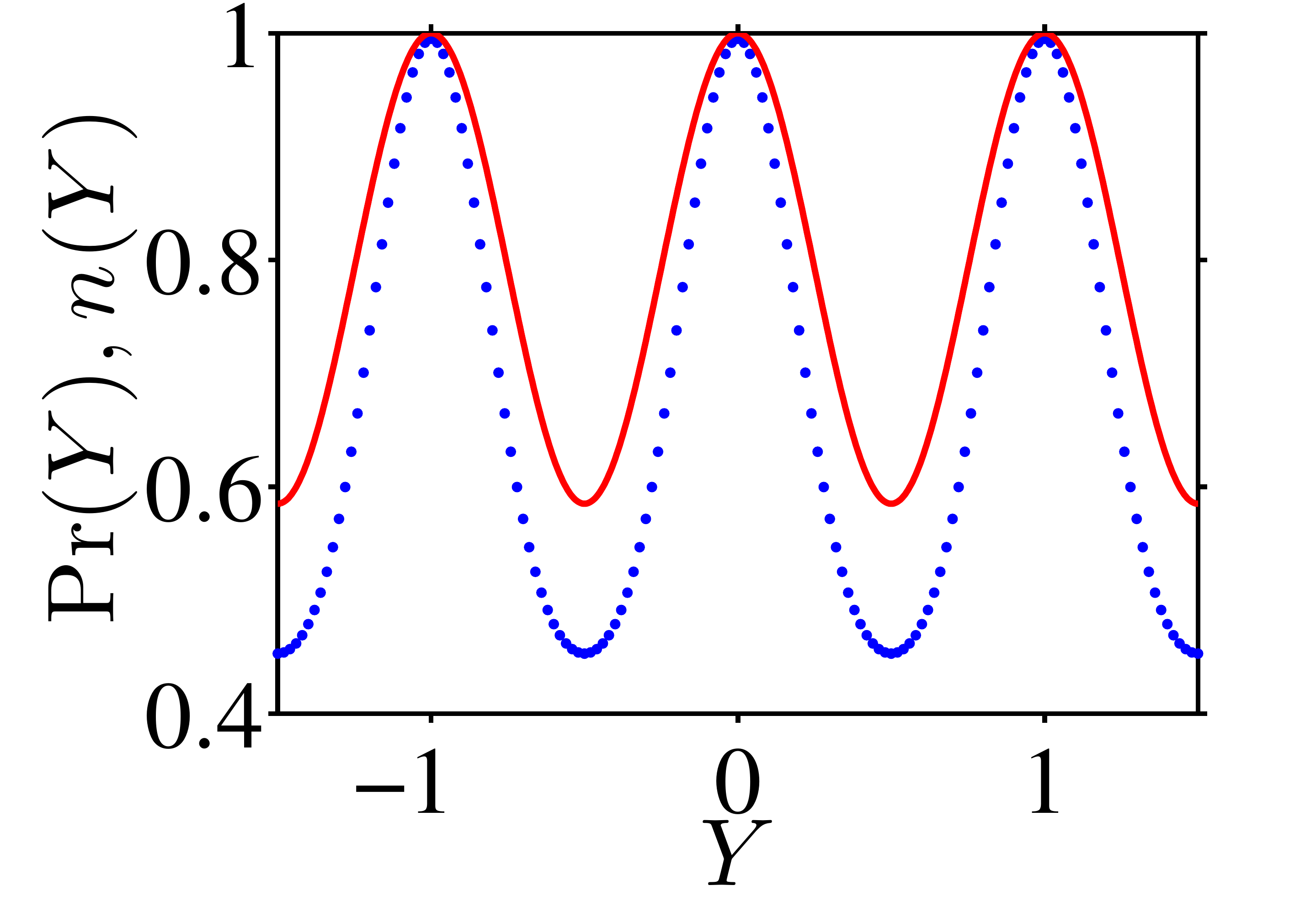}
\caption{The dotted line is ${\rm Pr(Y)}$ for $g=1$ and $\Theta=0.28\pi$, and the solid curve is the dipole density variation $n(Y)=n_0+n_1\cos(q_c Y)$ along the $y$ direction. Both quantities are normalised to their respective maximum values. The lengths are in units of $2\pi/q_c$.}
\label{psiy}
\end{figure}
\end{center}

To understand the coexistence of stripe and superfluid orders, we examine the bare particle pair correlation function
$\calC_P(\bk,-\bk) =  \left |\la \hat c_{\bk}\hat c_{-\bk} \ra \right |^2 $
which is  plotted in Fig.~\ref{pair-corr} (left) for $g=1$ and $\Theta=0.28\pi$.
It clearly shows  that  pairing is concentrated in the gapless regions of the underlying Fermi surface for the stripe phase. Consequently, it does not affect the particle distribution in the gapped region which is responsible for the stripe formation. We analyse this further by determining the pair wave function in real space $\psi_{pair}(\brho,\brho') \equiv \la \hat\psi(\brho)\hat \psi(\brho')\ra$, where $\hat \psi(\brho)$ is the field operator of the dipoles. In Fig.~\ref{pair-corr} (right) we show $|\psi_{pair}(\brho,\brho')|^2$ for a Cooper pair with the centre of mass at the origin of the coordinates. The $p$-wave nature of the pairing is clearly visible with $|\psi_{pair}(\brho,-\brho)|^2$ strongly peaked along the $x$-axis, 
where the dipole-dipole interaction is most attractive. In addition, we plot in Fig.~\ref{psiy} the relative probability density of finding the centre-of-mass of a Cooper pair at a specific location. This is given by ${\rm Pr}(Y)\equiv \int d(\brho-\brho') |\psi_{pair}(\brho,\brho')|^2$, which depends only on the $y$-coordinate of the Cooper pair centre-of-mass due to the translational symmetry in the $x$ direction. 
We see that the probability density varies in phase with that of the density of the dipoles, such that the $p$-wave pairing has a maximum 
on the stripes and a minimum in between. Thus, 
 from the real space perspective, the stripe and superfluid orders coexist due to the anisotropy of the dipolar interaction. Namely, the repulsive 
 part induces the stripe formation while the attractive part induces pairing, resulting in Cooper pairs with $p$-wave symmetry along the stripes as illustrated in Fig.\ \ref{SetupFig}. In the limit of strong interaction, the density between the stripes presumably vanishes and the stripes well separate. This raises 
  the interesting possibility of realising an array of 1D $p$-wave superconductors which have topological properties~\cite{Alicea}. The study 
  of this strong coupling limit is beyond the scope of the present paper. 

 We argued earlier that $\Theta_s\simeq \arcsin(2/3)$ is a good approximation for the boundary  
 separating the stripe and the supersolid phases for $g> g_c(\Theta)$. We now obtain a more accurate result by varying the tilting angle and determining the critical angle
 $\Theta_s(g)$ below which the gap equation ceases to admit a finite solution. 
Such calculations can also be carried out for the normal Fermi liquid to superfluid transition. The overall phase boundary obtained this way is shown in Fig.~\ref{PhaseFig}. 
We see that our initial estimate is in fact remarkably accurate and the phase boundary has a rather weak dependence on the interaction strength $g$. This suggests that the onset of pairing is primarily determined by the degree of anisotropy of the dipolar potential. The identification of the supersolid region in the phase diagram bounded by this boundary, $g_c(\Theta)$ and the collapse line, and our elucidation of the nature of this phase, are the main results of this paper. 

Finally, we  discuss how the supersolid phase  can be detected in time-of-flight (TOF) experiments, which have  been used to probe 
a variety of phases and correlations~\cite{Greiner,Folling,Rom,Spielman}. As shown in Ref.~\cite{Block2}, 
the density wave order can be detected by measuring the correlation function
 $\calC_D(\bk,\bk+\bq_c) \equiv |\la\hat c^\dag_\bk\hat c_{\bk+\bq_c}\ra|^2$
  in TOF experiments.  Similarly the superfluid order can be detected by a measurement of the pair correlation function $\calC_P(\bk,-\bk)$, which can then be 
 compared  to theoretical results such as that shown in Fig.~\ref{pair-corr} (left). However, we need to bear in mind that in a standard experiment, the imaging system introduces a smoothening of the absorption images in the $xy$-plane, which can be modelled by convolution of the absorption density with a Gaussian~\cite{Folling,BruunAnti}. This  reduces the magnitude of the correlation peak considerably from the theoretical maximum value of approximately $1/4$ shown in Fig.~\ref{pair-corr}. Nevertheless, we expect that the TOF experiments can be used to detect the supersolid phase. 

In conclusion, we  demonstrate that a 2D gas of fermionic dipoles aligned by an external field allow for a coexistence of stripe and superfluid order in a large region of the zero temperature phase diagram. This occurs as a result of the anisotropic nature of the dipolar interaction, where the repulsive part drives the stripe formation, and the attractive part induces the formation of $p$-wave Cooper-pairs along the stripes. In momentum space, the existence of the 
supersolid phase can be understood from the fact that the stripe order renders the Fermi surface partially gapped, leaving  
gapless the regions most important for $p$-wave pairing. We finally discuss how the supersolid phase can be detected in TOF experiments. 
Our results point to several interesting future research directions. This includes realising an array of 1D topological superconductors
 in the limit of strong interaction, and investigating parallels to the high $T_c$ cuprates, where the co-existence of charge-density-wave order and superconductivity was recently observed~\cite{Hayward}. 
 
\begin{acknowledgements}
GMB would like to acknowledge the support of the Hartmann Foundation via grant A21352 and the Villum Foundation via grant VKR023163.
\end{acknowledgements}

\clearpage

\section{Supplemental Material}

\subsection{1: Hartree-Fock theory on the stripe phase and the extended zone scheme}
The mean-field Hamiltonian used to describe this phase is given by~\cite{Block2} 
\beq
\hat \calH_{MF} =\sum_{\bk}\e_\bk\hat c_\bk^\dag \hat c_\bk +\sum_\bk[h_\bk\hat c^\dag_{\bk+\bq_c}\hat c_\bk + h.c.],
\label{Hdwp}
\eeq
where $\bq_c = q_c \hat {\bf y}$, $\e_\bk$ is the single particle Hartree-Fock energy 
\beq
\e_\bk = \frac{k^2}{2m} + \frac{1}{A}\sum_{\bk'}[V(0)-V(\bk-\bk')]\la \hat c_{\bk'}^\dag \hat c_{\bk'}\ra
\label{ek}
\eeq
and $h_\bk$ is a real off-diagonal element defined by
\beq
h_\bk = \frac{1}{A}\sum_{\bk'}[V(\bq_c)-V(\bk-\bk')]\la \hat c^\dag_{\bk'}\hat c_{\bk'+\bq_c}\ra.
\label{hk}
\eeq
The inclusion of the second term in Eq.~(\ref{Hdwp}) accounts for the possibility of formation of the density wave along the $y$ direction.
The Hamiltonian in Eq.~(\ref{Hdwp}) resembles (although is not identical to) that of non-interacting particles in a potential periodic in the $y$ direction with periodicity $2\pi/q_c$. Consequently the quasiparticle eigenlevels  $\varepsilon_{j\bar\bk}$ of Eq.~(\ref{Hdwp}) exhibit a band-like structure along the $y$ direction of the wave vector, where $j=1,2,\cdots$ is the band index and $\bar\bk$ is restricted to the first Brillouin zone. The corresponding quasiparticle wave function $\psi_{j\bar\bk}(\brho)$ can be expressed as
$
\psi_{j\bar\bk}(\brho) =  \sum_\bG U_{j,\bar\bk+\bG}e^{i(\bar\bk + \bG)\cdot\brho}/{\sqrt{A}}
$, where $\bG = l \bq_c, l =0,\pm 1,\cdots$ is the reciprocal lattice vector and 
the expansion coefficients $U_{j,\bar\bk+\bG}$ are determined the Schr{\"o}dinger equation
\beq
\left (\e_{\bar\bk+\bG}-\varepsilon_{j\bar\bk} \right ) U_{j,\bar\bk+\bG}+\sum_{\bG'=\bG\pm\bq_c}h_{\bar\bk+\bG'} U_{n,\bar\bk+\bG'}=0.
\label{Matreq}
\eeq
Equation (\ref{Matreq}) is analogous to the Schr{\"o}dinger equation of a particle in a periodic lattice, where $h_{\bar\bk+\bG'}$ plays the role of  the Fourier components of a ``periodic potential". Unlike a true periodic potential, however, $h_{\bar\bk+\bG'}$ depends explicitly on $\bar\bk$ due to the inclusion of the exchange interaction. In terms of the quasiparticle wave function basis, the Hamiltonian in (\ref{Hdwp}) can now be brought into a diagonalised form
$
\hat \calH_{MF}=\sum_{j\bar\bk} \varepsilon_{j\bar\bk}\hat \gamma_{j\bar\bk}^\dag\hat \gamma_{j\bar\bk}
$
, where $\hat \gamma_{j\bar\bk} = \sum_\bG U_{j,\bar\bk+\bG}\hat c_{\bar\bk + \bG}$ is the annihilation operator of the quasiparticle. 
 The quasiparticle occupation number in the ground state 
 $
 N_{j\bar\bk}=\la\hat \gamma_{j\bar\bk}^\dag\hat \gamma_{j\bar\bk} \ra = \theta(\mu-\varepsilon_{j\bar\bk})
 $ 
 is specified by the chemical potential $\mu$ of the density wave phase, which in turn is determined by the density of the gas as
\beq
n_0 =\frac{1}{A}\sum_{j\bar\bk}\theta(\mu- \varepsilon_{j\bar\bk} ).
\label{rho0}
\eeq
The quasiparticle energy $\varepsilon_{j\bar\bk}$ and the expansion coefficients $U_{j,\bar\bk+\bG}$ are implicit functions of the Hartree-Fock elements $\e_\bk$ and $h_\bk$. Therefore these quantities as well as the chemical potential $\mu$ are determined self-consistently through Eqs.~(\ref{ek})-(\ref{rho0}). The reader is referred to Ref.~\cite{Block2} for a detailed account of their numerical calculation. 

It turns out that the effects of the off-diagonal terms in Eq.~(\ref{Hdwp}) to the quasiparticle dispersion are only perturbative, due to the fact the magnitudes of $h_\bk$ are generally small compared to the Fermi energy $\e_F^{0}$~\cite{Block2}. Consequently the quasiparticle dispersion does not in fact deviate significantly from the usual parabolic form except in regions close to the Brillouin zone boundaries where band gaps open up. It is thus meaningful to use  the ``extended zone scheme"~\cite{Ashcroft} instead of the ``reduced zone scheme" in labelling the quasiparticle energy levels. More specifically, each of the physical quantities associated with the single particle state $\psi_{j\bar\bk}(\brho)$ can be labelled by a single wave vector in the $j$-th Brillouin zone $\bk=\bk_j$, which is defined as 
\beq
\bk_j = \left \{ \begin{array}{ll}
\bar\bk + \frac{j}{2}\bq_c, &\quad-q_c/2<\bar\bk\cdot\hat \bq_c\le 0 \\
\bar\bk - \frac{j}{2}\bq_c, &\quad 0 <\bar\bk\cdot\hat \bq_c\le q_c/2\end{array} \right.
\eeq
for $j = 2, 4,\cdots$ and
\beq
\bk_j = \left \{ \begin{array}{ll}
\bar\bk - \frac{j-1}{2}\bq_c, &\quad-q_c/2<\bar\bk\cdot\hat \bq_c\le 0 \\
\bar\bk + \frac{j-1}{2}\bq_c, &\quad 0 <\bar\bk\cdot\hat \bq_c\le q_c/2\end{array} \right.
\eeq
for $j = 1,3,\cdots$. Likewise, the physical quantities associated with the time-reversal state  $\psi_{j,-\bar\bk}(\brho)$ can be labelled by the vector $-\bk$. The effective pairing interaction, given by  
\begin{gather}
\calV_{j'j}(\bar\bk',-\bar\bk) = \sum_{\bG\bG'\tilde\bG \tilde \bG'}\delta_{\bG-\bG',\tilde \bG'-\tilde\bG}U^*_{j',\bar\bk'+\bG'} U^*_{j',-\bar\bk'+\tilde\bG'}\nonumber\\
\times U_{j,-\bar\bk+\tilde\bG}U_{j,\bar\bk+\bG}V\left(\bar\bk-\bar\bk'+\bG-\bG'\right ), 
\label{Vmatrdw2}
\end{gather}
shall be denoted by $\calV(\bk,-\bk')$. These correspondences are made clear if one considers the limit of vanishing off-diagonal Hartree-Fock elements $h_\bk$. In this limit the Bloch state $\psi_{j\bar\bk}(\brho)$ simply approaches the plane wave state $e^{i\bk_n\cdot\brho}/\sqrt{A}$ and the effective pairing interaction $\calV(\bk,-\bk')$ approaches the bare interaction $V(\bk-\bk')$.

\subsection{2: The pairing symmetry}
Here we provide a rationale for the choice of a gap parameter $\Delta(\bk)$ with even parity with respect to $\phi$ as expressed in the main Letter. The gap equation (4) in the main Letter can be written as 
\beq
\Delta_\bk = -\frac{1}{2}\int\frac{d\bk'}{(2\pi)^2}\tilde \calV(\bk,-\bk')\Delta_{\bk'}\left (\frac{1}{\calE_{\bk'}} +\frac{\calP}{\mu-k^{\prime 2}/2m}\right ),
\label{geqdw1}
\eeq
where $\tilde \calV(\bk,-\bk')$ is the anti-symmetrized interaction matrix
\beq
\tilde \calV(\bk,-\bk') =\frac{1}{2}[\calV(\bk,-\bk')-\calV(\bk,\bk')].
\label{ascalV}
\eeq
It can be shown that the interaction matrix $\tilde \calV(\bk,-\bk')$ has the following expansion
\begin{align}
 \tilde \calV(\bk,-\bk') = \sideset{}{'}\sum_{n,n'=1}^\infty \left [\calV^{cc}_{nn'}(k,k')\cos n\phi\cos n'\phi' \right . \nn \\
 \left. +\calV^{ss}_{nn'}(k,k')\sin n\phi\sin n'\phi' \right ],
\label{tcVpwexp2}
\end{align}
where $\sum^\prime$ restricts the summation to odd indices, 
\begin{align}
\calV^{cc}_{nn'}(k,k') = \int_0^{2\pi}\frac{d\phi}{\pi} \int_0^{2\pi}\frac{d\phi'}{\pi} \cos n \phi \cos n' \phi' \calV(\bk,-\bk')
\label{calVcc}
\end{align}
and
\begin{align}
 \calV^{ss}_{nn'}(k,k') =\int_0^{2\pi}\frac{d\phi}{\pi} \int_0^{\pi}\frac{d\phi'}{\pi}  \sin n \phi \sin n' \phi' \calV(\bk,-\bk').
\label{calVss}
\end{align}
The sine and cosine terms in the expansion (\ref{tcVpwexp2}) are not coupled and, as a consequence, the gap equation (\ref{geqdw1}) admits solutions with either even or odd parity with respect to the $\phi$ variable. For even solutions only the first part of the potential in (\ref{tcVpwexp2}) contributes to the integral in Eq.~(\ref{geqdw1}) and for odd solutions only the second part does. 
In order to see which part of the potential favours Cooper pairing, we express $\calV^{cc}_{nn'}(k,k')$ and $\calV^{ss}_{nn'}(k,k') $ in terms of the interaction potential in real space $V(\brho)=V(\rho,\theta)$. This can be done by approximating $\calV(\bk,-\bk')$ by the bare interaction matrix
\begin{align}
V(\bk-\bk')=\int d\brho V(\rho,\theta) e^{-ik\rho\cos(\phi-\theta)}e^{-ik'\rho\cos(\phi'-\theta)}
\label{V2Df3}
\end{align}
in Eqs.~(\ref{calVcc}) and (\ref{calVss}). Using the expansion
$
e^{ix\cos\theta} = \sum_{n=-\infty}^\infty i^n J_n(x)e^{in\theta}
$,
where $J_n(x)$ is the Bessel function of the first kind, and performing the integrals with respect to $\phi$ and $\phi'$, we find
\begin{align}
\calV^{cc}_{nn'}(k,k')& \simeq 4i^{n'-n}\int_0^\infty \rho d\rho\int_0^{2\pi}d \theta  \nn \\
&\times \cos n \theta \cos n' \theta J_n(k\rho) J_{n'}(k'\rho) V(\rho,\theta)
\label{Vccreal}
\end{align}
and
\begin{align}
 \calV^{ss}_{nn'}(k,k') &\simeq 4i^{n'-n}\int_0^\infty \rho d\rho\int_0^{2\pi}d \theta \nn \\
 &\times \sin n\theta \sin n'\theta  J_n(k\rho) J_{n'}(k'\rho) V(\rho,\theta).
 \label{Vssreal}
\end{align}
From these expressions we see that $\calV^{cc}_{nn'}(k,k')$ mostly samples the attractive part of the potential in the real space (the sliver around the $x$ axis) while $\calV^{ss}_{nn'}(k,k')$ mostly samples the repulsive part. Let us take the the first diagonal elements $\calV^{cc}_{11}(k,k')$ 
and $\calV^{ss}_{11}(k,k')$
for example. These are in fact the most dominant matrix elements for $\calV^{cc}_{nn'}(k,k')$ and $\calV^{ss}_{nn'}(k,k')$ respectively. As the attractive sliver of the potential $V(\brho)$ expands from the $x$ axis with an increasing tilting angle $\Theta$, $\calV^{cc}_{11}(k,k')$ can potentially become negative due to the fact the attractive part of potential is more significantly weighted in the integral. The matrix element $\calV^{ss}_{11}(k,k')$, on the other hand, remains positive for all tilting angles. This analysis motivates us to look for solutions to the gap equation with even parity with respect to $\phi$.

\subsection{3: Comparisons between $\calV^{cc}_{nn'}(k,k')$ and $V^{cc}_{nn'}(k,k')$}
\begin{center}
\begin{figure}[ht]
\includegraphics[width=0.9\linewidth]{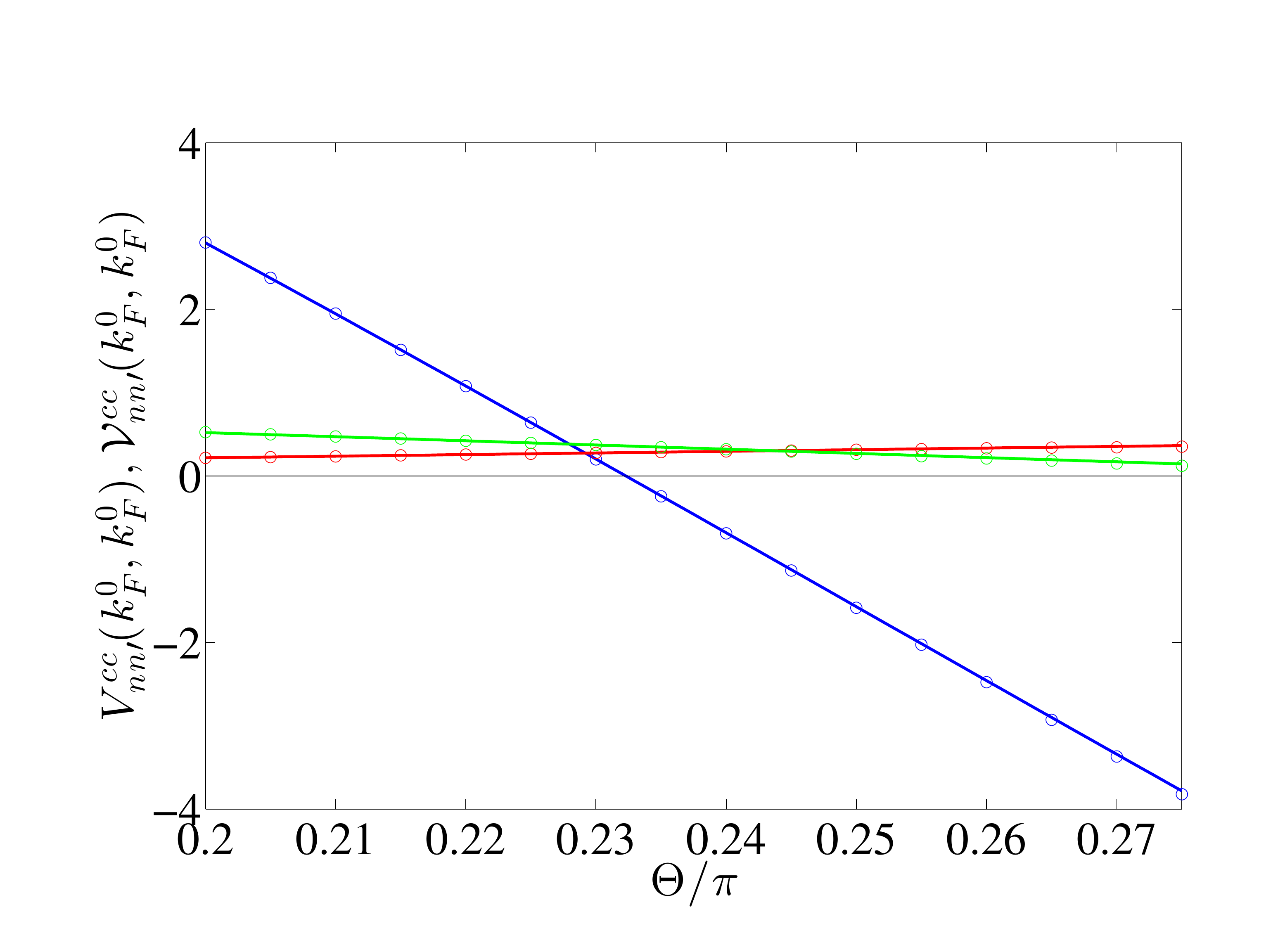}
\caption{The matrix elements (in units of $g/m$) $\calV^{cc}_{11}(k^0_F,k^0_F)$ (blue circle), $\calV^{cc}_{13}(k^0_F,k^0_F)$ (red circle)  and $\calV^{cc}_{33}(k^0_F,k^0_F)$ (green circle)  as a function of $\Theta$ for $g= 0.95$. The solid lines are $V^{cc}_{11}(k^0_F,k^0_F)$ (blue), $V^{cc}_{13}(k^0_F,k^0_F)$ (red) and $V^{cc}_{33}(k^0_F,k^0_F)$ (green)  respectively.}
\label{Vnn_comp}
\end{figure}
\end{center}

\begin{center}
\begin{figure}[ht]
\includegraphics[width=0.9\linewidth]{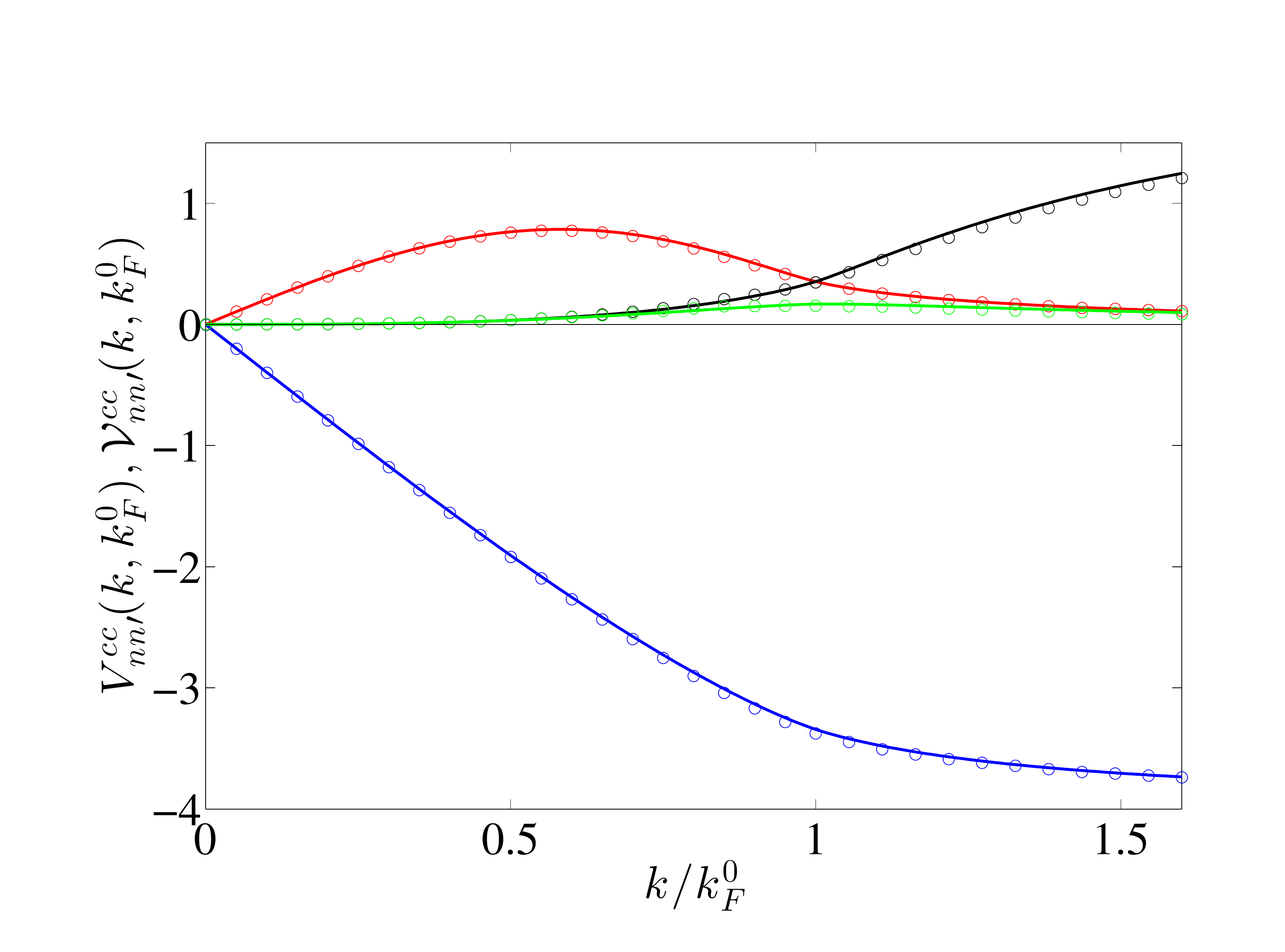}
\caption{The matrix elements (in units of $g/m$) $\calV^{cc}_{11}(k,k^0_F)$ (blue circle), $\calV^{cc}_{13}(k,k^0_F)$ (red circle), $\calV^{cc}_{31}(k,k^0_F)$ (black circle) and $\calV^{cc}_{33}(k,k^0_F)$ (green circle) as a function of $k/k_F^0$ for $g= 0.95$ and $\Theta=0.27\pi$. The solid lines are $V^{cc}_{11}(k,k^0_F)$ (blue), $V^{cc}_{13}(k,k^0_F)$ (red), $V^{cc}_{31}(k,k^0_F)$ (black) and $V^{cc}_{33}(k,k^0_F)$ (green) respectively.}
\label{Vnn_comp2}
\end{figure}
\end{center}

The Fourier components $V^{cc}_{nn'}(k,k')$ (which are defined by Eq.~(\ref{calVcc}) with $\calV(\bk,-\bk')$ replaced by $V(\bk-\bk')$) for the bare interaction can be evaluated analytically. 
Using Eq.~(2) of the main Letter in Eq.~(\ref{calVcc}) we find that the only non-vanishing matrix elements are those whose indices differ by 0 or $\pm 2$. That is, the matrix $ V^{cc}_{nn'}(k,k')$ has the following tridiagonal structure
\begin{align}
V^{cc}_{nn'}(k,k')& = \delta_{n,n'} V^{cc}_{nn}(k,k') +\delta_{n,n'+2} V^{cc}_{n+2,n}(k,k') \nn \\
&\quad + \delta_{n,n'-2} V^{cc}_{n,n+2}(k,k').
\label{Vccstruc}
\end{align}
 For odd $n$ we find
\begin{align}
V^{cc}_{nn}(k,k') 
= &\frac{3\pi g}{m}\frac{kk'}{k^0_F(k+k')}\frac{1}{n} [I_{n-1}(k,k')-I_{n+1}(k,k')]\nn \\
&\times\left [1-\left(\frac{3}{2}+\delta_{n,1}\frac{3}{4}\right )\sin^2\Theta\right ],
\label{Vccd}
\end{align}
\begin{align}
&V^{cc}_{n+2,n}(k,k') 
= -\frac{ 3\pi g}{2m}\frac{kk'}{k^0_F(k+k')} \nn \\
&\times\left [ \frac{k}{k'}I_{n}(k,k')+\frac{k'}{k}I_{n+2}(k,k') +2 I_{n+1}(k,k') \right]\sin^2\Theta,
\label{Vccud}
\end{align}
and
\begin{align}
&V^{cc}_{n,n+2}(k,k')
= -\frac{ 3\pi g}{2m}\frac{kk'}{k^0_F(k+k')} \nn \\
&\times\left [\frac{k}{k'} I_{n+2}(k,k')+\frac{k'}{k}I_{n}(k,k') +2I_{n+1}(k,k') \right ]\sin^2\Theta,
\label{Vccld}
\end{align}
where
\beq
I_m(k,k') = \int_0^{\frac{\pi}{2}}d\phi \frac{\cos2m\phi}{\sqrt{1-x^2\sin^2\phi}}
\eeq
with $x \equiv \sqrt{4kk'/(k+k')^2}$. The integral $I_m(k,k')$ can generally be expressed in terms of the complete elliptic integrals. As a few examples, $I_m(k,k')$ for $m=0,1,2$ are shown below as
\beq
I_0(k,k') = K(x),
\eeq
\beq
I_1(k,k') = \frac{1}{x^2}[K(x)x^2-2K(x)+2E(x)],
\eeq
and
\begin{align}
I_2(k,k') &= \frac{1}{3x^4}\left [3K(x)x^4-16K(x)x^2 \right. \nn \\
&\left.+8E(x)x^2+16K(x)-16E(x)\right ],
\end{align}
where $K(x)$ and $E(x)$ are the complete elliptic integrals of the first and second kind respectively.  
As a useful gauge of the relative importance of the matrix elements for various $n$, we consider the special case of $k = k'$. In this case the expressions in (\ref{Vccd})-(\ref{Vccld}) simplify and we find  
\begin{align}
V^{cc}_{nn}(k,k) =\frac{12\pi g}{(4n^2-1)m} \frac {k}{k^0_F}\left[1-\left (\frac{3}{2} +\delta_{n,1}\frac{3}{4}\right )\sin^2\Theta \right ]
\label{Vnnkk}
\end{align}
and
\begin{align}
V^{cc}_{n+2,n}(k,k) &= V^{cc}_{n,n+2}(k,k) \nn \\
&=\frac{3\pi g}{(2n+1)(2n+3)m}\frac{k}{k^0_F} \sin^2\Theta.
\label{Vn2nkk}
\end{align}
We see that the magnitudes of these matrix elements decreases rapidly as $1/n^2$. 
The Fourier components $\calV^{cc}_{nn'}(k,k')$ for the quasiparticles are calculated numerically using Eq.~(\ref{calVcc}). We find that the formation of density wave has minimal effects on the effective paring interaction, namely $\calV^{cc}_{nn'}(k,k')$ agrees very well with $V^{cc}_{nn'}(k,k')$. (see Figs.~\ref{Vnn_comp}-\ref{Vnn_comp2} as an example) for a wide range of $g$ and $\Theta$. 

\end{document}